\documentclass{article}

\usepackage[preprint]{neurips_2024}
\usepackage{amsmath}

\usepackage[utf8]{inputenc} 
\usepackage[T1]{fontenc}    
\usepackage{hyperref}       
\usepackage{url}            
\usepackage{booktabs}       
\usepackage{amsfonts}       
\usepackage{nicefrac}       
\usepackage{microtype}      
\usepackage{xcolor}         
\usepackage{graphicx}
\usepackage{multirow}

\title{Bridging Auditory Perception and Language Comprehension through MEG-Driven Encoding Models}

\author{Matteo Ciferri \\
University of Rome, Tor Vergata\\
Department of Biomedicine and Prevention\\
\texttt{matteo.ciferri@students.uniroma2.eu} \\
\And
Matteo Ferrante \\
University of Rome, Tor Vergata\\
Department of Biomedicine and Prevention\\
\texttt{matteo.ferrante@uniroma2.it} \\
\And
Nicola Toschi \\
University of Rome, Tor Vergata\\
Department of Biomedicine and Prevention\\
A.A. Martinos Center for Biomedical Imaging \\
Harvard Medical School/MGH, Boston (US) \\
}

\begin{document}

\maketitle

\begin{abstract}
Understanding the neural mechanisms behind auditory and linguistic processing is key to advancing cognitive neuroscience. In this study, we use Magnetoencephalography (MEG) data to analyze brain responses to spoken language stimuli. We develop two distinct encoding models: an audio-to-MEG encoder, which uses time-frequency decompositions (TFD) and wav2vec2 latent space representations, and a text-to-MEG encoder, which leverages CLIP and GPT-2 embeddings. 
Both models successfully predict neural activity, demonstrating significant correlations between estimated and observed MEG signals. However, the text-to-MEG model outperforms the audio-based model, achieving higher Pearson Correlation (PC) score.
Spatially, we identify that auditory-based embeddings (TFD and wav2vec2) predominantly activate lateral temporal regions, which are responsible for primary auditory processing and the integration of auditory signals. In contrast, textual embeddings (CLIP and GPT-2) primarily engage the frontal cortex, particularly Broca’s area, which is associated with higher-order language processing, including semantic integration and language production, especially in the 8–30 Hz frequency range. The strong involvement of these regions suggests that auditory stimuli are processed through more direct sensory pathways, while linguistic information is encoded via networks that integrate meaning and cognitive control. Our results reveal distinct neural pathways for auditory and linguistic information processing, with higher encoding accuracy for text representations in the frontal regions. These insights refine our understanding of the brain's functional architecture in processing auditory and textual information, offering quantitative advancements in the modelling of neural responses to complex language stimuli.
\end{abstract}

\section{Introduction}
\begin{figure}[ht]
    \centering
    \includegraphics[width=0.99\linewidth]{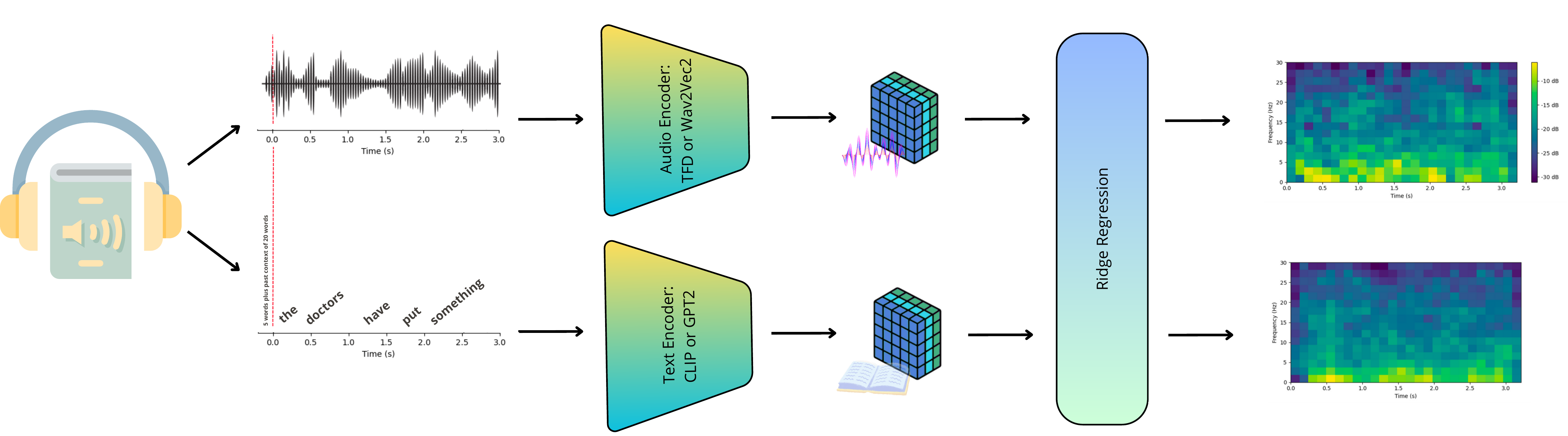}
    \caption{Schematic representation of the encoding pipelines. The top part of the figure is structured as follows: Left: initial input stimulus (audio). Centre: two different encoders individually process the auditory stimulus to generate embeddings. The first encoder uses time-frequency decompositions to extract features, while the second encoder uses the wav2vec2 library to convert the audio data into latent representations. Right: prediction of MEG time-frequency decompositions (e.g., spectrogram) using ridge regression with the embeddings as dependent variables. For text stimuli (bottom part of the figure), the same pipeline is adapted with CLIP and GPT-2 models as encoders, which generate embeddings that capture the semantic content and contextual information of the text. This setup allows us to compare and analyze how different types of stimuli are represented and processed in the brain.}
    \label{fig:model}
\end{figure}

In recent years, the field of computational neuroscience has seen significant advancements in understanding how the brain processes language. While much of the existing research in brain encoding and decoding \citep{goldstein_shared_2022, tang_semantic_2023} relies on functional Magnetic Resonance Imaging (fMRI) data, this modality is somewhat limited, amongst other factors, by its low temporal resolution. 
In contrast, the temporal resolution offered by Magnetoencephalography (MEG), despite other limitations (e.g. lower sensitivity in deep brain structures), could provide a more detailed and dynamic insight into neural mechanisms underlying language comprehension and generation. In this work, we aimed to develop encoding models to advance our understanding of language processing through the lens of MEG data. An encoding model is a computational framework designed to map input stimuli to corresponding (i.e. elicited by the corresponding stimulus) neural activity. Here, we develop audio-to-MEG encoders using two types of representations for audio data, i.e. time-frequency decompositions derived from Short-time Fourier Transform (STFT) \citep{GriffinLim1984}, and latent representations generated by the wav2vec2 library \citep{baevski2020wav2vec}. Additionally, we built text-to-MEG encoders that incorporate embeddings from the Contrastive Language-Image Pretraining (CLIP) model \citep{radford2021learning} or GPT-2 \citep{radford2019language} and compared the encoding performance between all pipelines (Figure~\ref{fig:model}). This comparison was performed with the goal of gaining insight into the neural processes involved in auditory and linguistic perception and advancing the computational strategies used for interpreting complex neural signals. 

\section{Related Work}
So far, research in brain encoding for speech and language processing has primarily used functional Magnetic Resonance Imaging (fMRI) \citep{Huth2012-od, antonello_scaling_2023, caucheteux_evidence_2023}. These studies have contributed to the development of both linear and nonlinear models that map stimuli to brain activity from fMRI signals. Previous work focused on e.g. enhancements in network scaling and uncovering associations between brain activity and specific auditory and semantic processing tasks \citep{caucheteux_brains_2022}. However, limitations in the temporal resolution of fMRI, which are particularly relevant in speech decoding due to the high-frequency content of the stimuli, have led researchers to explore MEG data collected during exposure to auditory stimuli. On the encoding side, \cite{oota2023meg} developed a model using Bidirectional Encoder Representations from Transformers (BERT) contextual embeddings \citep{devlin2018bert} to predict MEG signals. In terms of decoding, one paper \citep{defossez_decoding_2023} successfully reconstructed audio from MEG signals through contrastive learning which was based on aligning signals with the latent space generated by the wav2vec2 library \citep{baevski2020wav2vec}. These efforts demonstrate the potential of using MEG data to reconstruct the stimulus that has generated it. 
Our study builds upon these developments, aiming to augment encoding models by creating mappings of how the brain processes semantic and speech information.

\section{Material and Methods}
In this section, we outline the proposed method and the dataset we used. The data is publicly available and can be accessed at \url{https://osf.io/ag3kj/}. Code available at \url{https://github.com/neoayanami/meg-encoding}.

\subsection{Data}
We used data from the MEG-MASC dataset \citep{Gwilliams2023}, specifically selecting 8 subjects as in the study by \cite{oota2023meg}. The dataset includes recordings from 208 MEG sensors as the subjects listened to a series of naturalistic spoken stories, selected from the Open American National Corpus, namely “\textit{Cable Spool Boy}”, “\textit{LW1}”, “\textit{Black willow}”, and “\textit{Easy money}”. For pre-processing the raw MEG data, we employed the \textit{MNE-Python} library \citep{appelhoff2019mne}, which involved a) bandpass filtering  (0.5-30.0 Hz) \citep{marzetti2013frequency} b) segmentation into windows (length = 3 s) which begin in correspondence with a word (stimulus) onset, and typically encompass approximately 5 words; c) window-wise baseline correction using 200 ms of signal taken immediately before the stimulus to minimize noise from non-task-related variations (the mean signal across the baseline period is subtracted from all time points within the epoch); d) channel-wise clipping of amplitude signals between the fifth and ninety-fifth percentile. Also, the audio and MEG data were originally sampled at 16000 Hz and 1000 Hz, respectively. Given the necessity of temporal alignment between each audio window and the corresponding MEG segment, pre-processing resulted in a collection of 48000 time points for the audio signals  (\(3 s \times 16000 Hz \)) and of 3000 time points (\(3 s \times 1000 Hz \), baseline-corrected) for MEG each channel/sensor. From this point on with the 3-second windows term we will also include the baseline period in the case of MEG signal.

\subsection{Audio Encoding Models}
\textcolor{black}{We used time-frequency decompositions (TFDs) as a unified representation for both the input (speech signal) and the output (MEG signal). These were computed using Short-Time Fourier Transform (STFT) applied to 3-second windows defined as described above.
The Short-Time Fourier Transform (STFT) was applied to 3-second windows, as previously defined. The STFT parameters, including the number of Fast Fourier Transform points (n-FFT) and the overlap between frames (hop length), were adjusted to ensure temporal alignment between the MEG and speech signals. This setup produced a consistent representation of both signals, with each 3-second window divided into 26 time frames.}

Also, in order to apply the pre-trained wav2vec2 model, we processed inputs comprising the 3-second audio stimuli, each sampled at a frequency of 16000 Hz, obtaining 48000 time points. Single windows derived from the audio signal are then encoded into a \( (149, 768) \) embedding matrix from the last hidden layer. \\

\subsection{Text Encoding Models}
We designed an approach that aimed to associate each MEG window with the corresponding phrase, thus forming word sequences that incorporated the linguistic context of the latter. For each sequence, we included the 20 words preceding (i.e. past context) the stimulus (the word at onset) and the 5 words following the stimulus (which coincided with the 3-second MEG window under investigation). In order to encode the 25 words, we used the tokenizer from the pre-trained CLIP model to transform sentences into a format suitable for the encoding model. We accommodated sentence structure and additional linguistic elements by including padding as well as beginning-of-stream (BOS), and end-of-stream (EOS) tokens. Each sentence is therefore represented by a \( (33, 512) \) matrix, derived from the final hidden layer of the CLIP encoder. The first dimension corresponds to the encoded sentence, comprising both tokens derived from words and padding tokens, and the second is the embedding dimension. 
We also used the GPT-2 model for text analysis, using a pipeline similar to the one described above. A key distinction between GPT-2 and CLIP text encoders, however, lies in the dimensional structure of their embeddings. Specifically, the GPT-2 model has a larger size in its last hidden layer, with a feature vector of length 768. This larger embedding space allows for a potentially richer and more nuanced representation of textual data.

\subsection{Reconstruction from Audio and Text Embeddings}

Audio and text embeddings were then used to predict brain responses in form of TFDs. As in e.g. \cite{oota2023meg}, the last module of all encoding models was defined as a ridge regression layer. The ridge regression objective function for the latent representations of stimuli is \( f(X_s) = \min \lVert Y_b - X_s W_s \rVert^2_F + \lambda \lVert W_s \rVert^2_F \). Here, \( X_s\) represents the input embeddings, \( W_s \in \mathbb{R}^{F_s \times L}\) are the learnable weights, with \( F_s\) denoting the stimulus embeddings (which can be derived from either audio or textual data) and \( L\) the number of MEG sensors. The sample stimulus \( s \in \mathbb{R}^{F_s}\), \( \lVert . \rVert_F\) indicates the Frobenius norm, and \( \lambda>0\) is the regularization weight, a tunable hyper-parameter. \\
The dataset underwent subject-wise 70/30 train-test splitting. Training procedures involved cross-validations on the training set. Optimization of the parameter \( \lambda\), was conducted by exploring different values (1, 10, 500, 5000). Following cross-validation, the model was retrained on the entire training split using the best-performing hyperparameter (5000).

We evaluated the reconstructed brain responses between 0.5-30 Hz (i.e. the "complete" spectrum) as well as for individual frequency bands i.e. delta, theta, alpha, and beta \citep{ABHANG201619}. Delta frequencies are typically between 0.5 and 4 Hz, often associated with deep sleep or states of unconsciousness. Theta frequencies are generally between 4 and 8 Hz, associated with states between wakefulness and sleep. Alpha frequencies are typically from 8 to 12 Hz referring to relaxed, calm states while awake, and finally, beta frequencies between 12 and 30 Hz are linked to active, busy, or anxious thinking and concentration \citep{LOPESDASILVA20131112}. 

\subsection{Evaluation and Statistical validation}

In order to evaluate the quality of the reconstructed TFDs, we computed the Pearson Correlation \citep{siems2016measuring} and the coefficient of determination R² among every pair of (flattened) real and predicted TFDs. This operation was conducted for each MEG sensor and subject, which was followed by averaging across subjects to obtain an average, sensor-wise evaluation. We also averaged across sensors for global performance evaluation. 

In addition, in order to control for potential spurious results, we constructed a distribution for the null hypothesis that the PC values between real and predicted TFDs are zero, by randomly permuting the reconstructed, flattened TFDs, across TFD elements and within each subject and channel. We used 100 repetitions to build an empirical distribution of PC values between the "real" flattened TFDs and the latter 100 randomly permuted TFDs \citep{tang_semantic_2023}, resulting in a \( (8, 208, 100) \) matrix. 
The position of each  "true" PC within the corresponding empirical distribution was used to calculate a P-value, as well as to Z-transform all PC values. Subsequently, Z-scores and P-values were visualized channel-wise (averaging across subjects) and subject-wise (averaging across channels). P \(<\) 0.05 was considered statistically significant.

\section{Results}

\subsection{Evaluation of Correlation Metrics}
\begin{figure}[ht]
    \centering
    \includegraphics[width=0.95\linewidth]{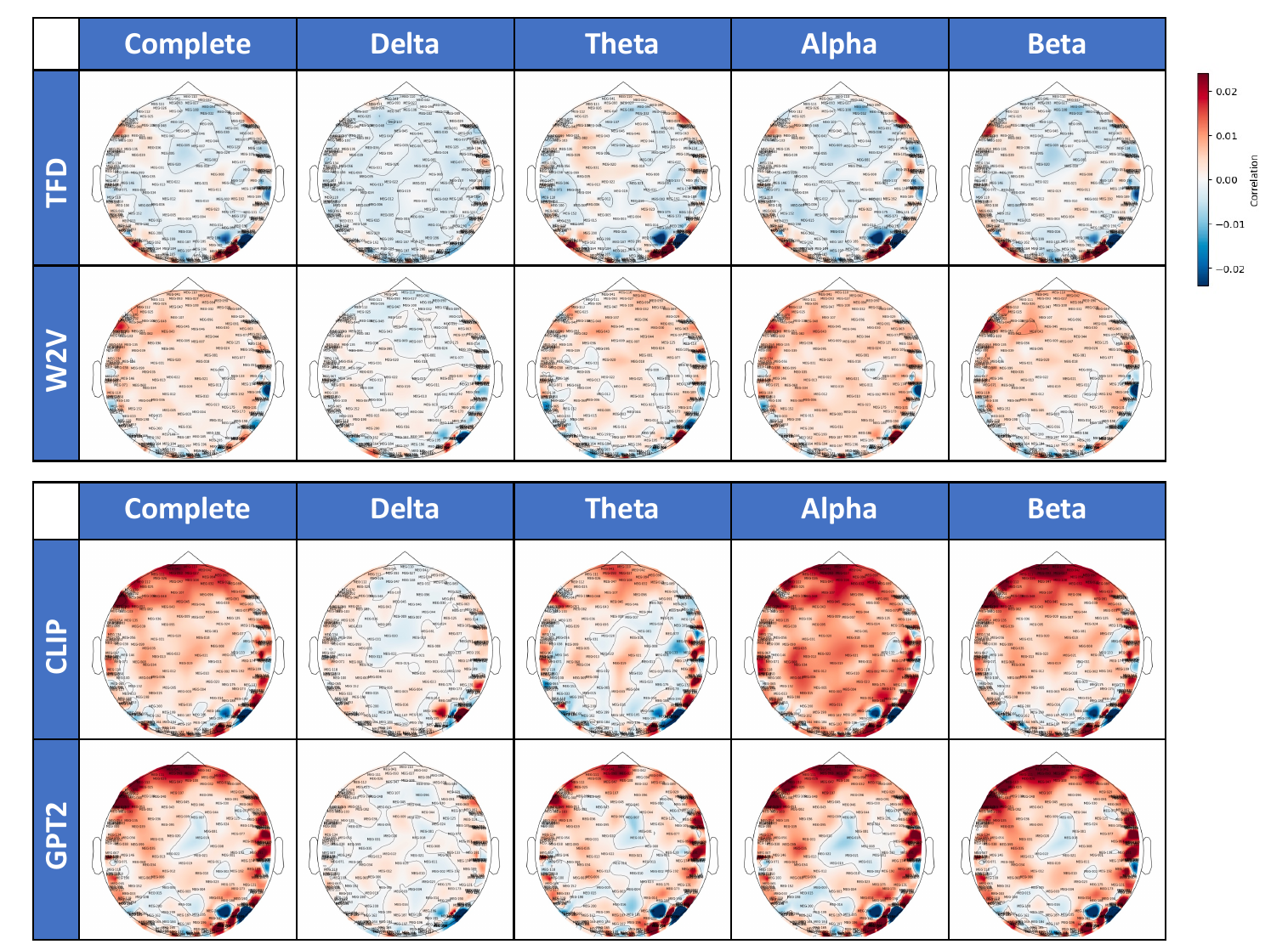}
    \caption{Pearson Correlation topography maps (subject-wise average), visualizing the performance of all encoding strategies (TFD, wav2vec2, CLIP, GPT-2) performance for every sensor and frequency band, as well as the full spectrum case ("complete"). In the case of audio encoders, high values of correlation occur in lateral brain areas, while textual models exhibit significant performance also in frontal regions. The performance varies notably across frequency bands, particularly in lower frequencies, which are typically associated with states of rest or sleep rather than concentration and cognitive processing.}
    \label{fig:topomaps}
\end{figure}

Figure~\ref{fig:topomaps} shows the anatomical distribution of Pearson Correlation scores for each sensor location and frequency band after subject-wise averaging. The highest scores were observed predominantly in lateral brain areas for the audio encoders and also in frontal regions for the textual models. 

\begin{table}[ht]
\centering
\caption{Comparative Pearson Correlation and R² results from audio and text encoders.}
\label{tab:combinedmodelresults}
\vspace{6pt}
\begin{tabular}{lcccccc}
\hline
\multirow{2}{*}{\textbf{Band}} & \multirow{2}{*}{\textbf{Input}} & \multirow{2}{*}{\textbf{Model}} & \multicolumn{2}{c}{\textbf{PC \scriptsize{($10^{-3}$})}} & \multicolumn{2}{c}{\textbf{R2 \scriptsize{($10^{-4}$})}} \\ \cline{4-7}
& & & \textbf{mean} & \textbf{std.} & \textbf{mean} & \textbf{std.} \\ \hline
\multirow{4}{*}{Complete} & \multirow{2}{*}{Audio} & TFD & 2.11 & 2.41 & 0.28 & 0.46 \\
& & wav2vec2 & 2.56 & 2.08 & 0.26 & 0.30 \\
& \multirow{2}{*}{Text} & CLIP & 5.60 & 4.80 & 1.87 & 3.19 \\
& & GPT-2 & 6.11 & 5.38 & 1.75 & 2.92 \\ \hline
\multirow{4}{*}{Delta} & \multirow{2}{*}{Audio} & TFD & 0.60 & 1.65 & 0.04 & 0.11 \\
& & wav2vec2 & 0.35 & 1.35 & 0.02 & 0.09 \\
& \multirow{2}{*}{Text} & CLIP & 1.31 & 1.77 & 0.10 & 0.15 \\
& & GPT-2 & 0.10 & 1.67 & 0.07 & 0.14 \\ \hline
\multirow{4}{*}{Theta} & \multirow{2}{*}{Audio} & TFD & 1.98 & 2.71 & 0.29 & 0.48 \\
& & wav2vec2 & 1.63 & 2.27 & 0.18 & 0.27 \\
& \multirow{2}{*}{Text} & CLIP & 1.52 & 4.62 & 0.64 & 1.56 \\
& & GPT-2 & 4.38 & 4.86 & 1.05 & 1.84 \\ \hline
\multirow{4}{*}{Alpha} & \multirow{2}{*}{Audio} & TFD & 2.03 & 3.08 & 0.36 & 0.61 \\
& & wav2vec2 & 3.85 & 2.44 & 0.53 & 0.49 \\
& \multirow{2}{*}{Text} & CLIP & 9.12 & 5.11 & 2.96 & 3.12 \\
& & GPT-2 & 6.65 & 5.95 & 2.20 & 3.12 \\ \hline
\multirow{4}{*}{Beta} & \multirow{2}{*}{Audio} & TFD & 2.33 & 2.83 & 0.40 & 0.66 \\
& & wav2vec2 & 2.84 & 2.57 & 0.35 & 0.46 \\
& \multirow{2}{*}{Text} & CLIP & 6.29 & 6.28 & 2.75 & 5.51 \\
& & GPT-2 & 7.45 & 6.61 & 2.70 & 3.84 \\ \hline
\end{tabular}
\end{table}

Table~\ref{tab:combinedmodelresults} shows an overview of all our results. We averaged the R² scores and the PC across all sensors and frequency bands. We observed that using textual embeddings for MEG encoding resulted in higher performance as compared to using speech embeddings. However, the latter still resulted in inhomogeneous performance distributions across brain regions, potentially offering physiological insight.

\subsection{Statistical Validation}
\begin{figure}[ht]
    \centering
    \includegraphics[width=.999\linewidth]{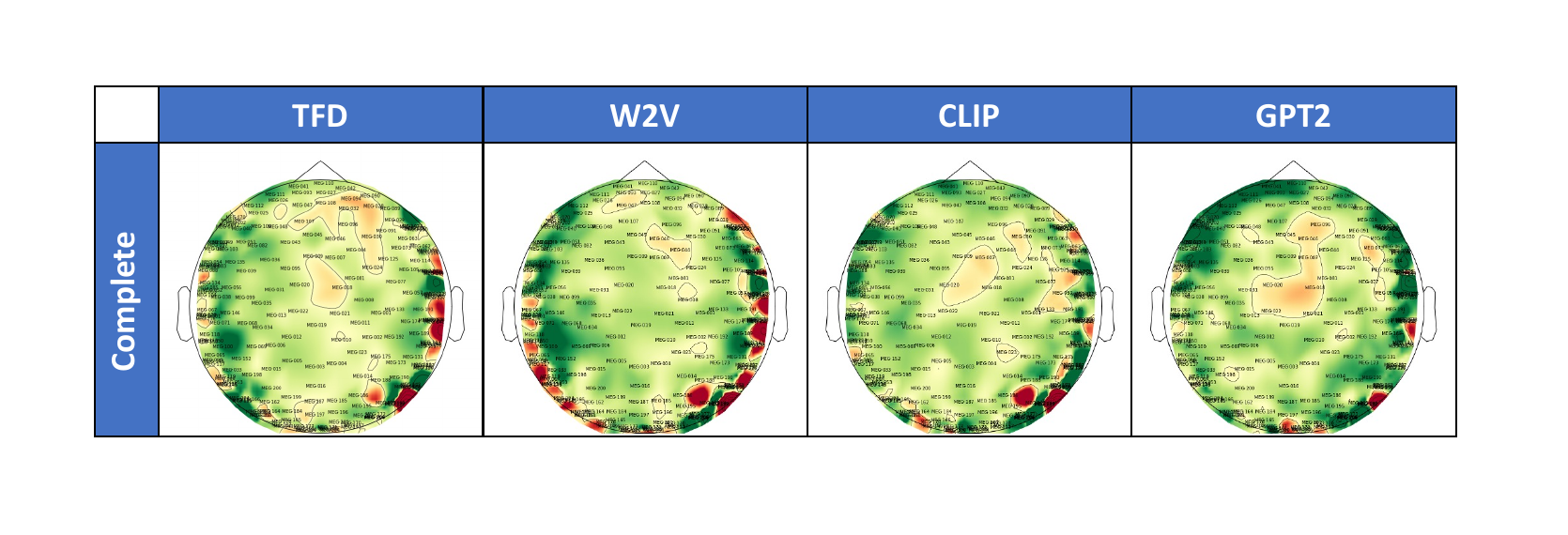}
    \caption{Topographic maps of PC Z-scores for the entire frequency range (complete, 0-30 Hz) across four different models (TFD, wav2vec2, CLIP, GPT-2). Each map illustrates the spatial distribution of Z-scores on the scalp, where Z-scores quantify the degree of association between the real and predicted MEG spectrograms. Higher Z-scores (indicated by green colour) represent a stronger positive association and suggest that the model's predictions are significantly different from what would be expected by chance. Conversely, negative Z-scores (marked by red colour) also indicate significant non-random associations, however in the opposite direction, highlighting regions where the model's predictions systematically deviate from the observed data. These maps provide a visual representation of the non-random patterns of brain activity predicted by each model, emphasizing the regions where the models' predictions are most robust and significant.}
    \label{fig:topomapsz}
\end{figure}

\begin{figure}[ht]
    \centering
    \includegraphics[width=.99\linewidth]{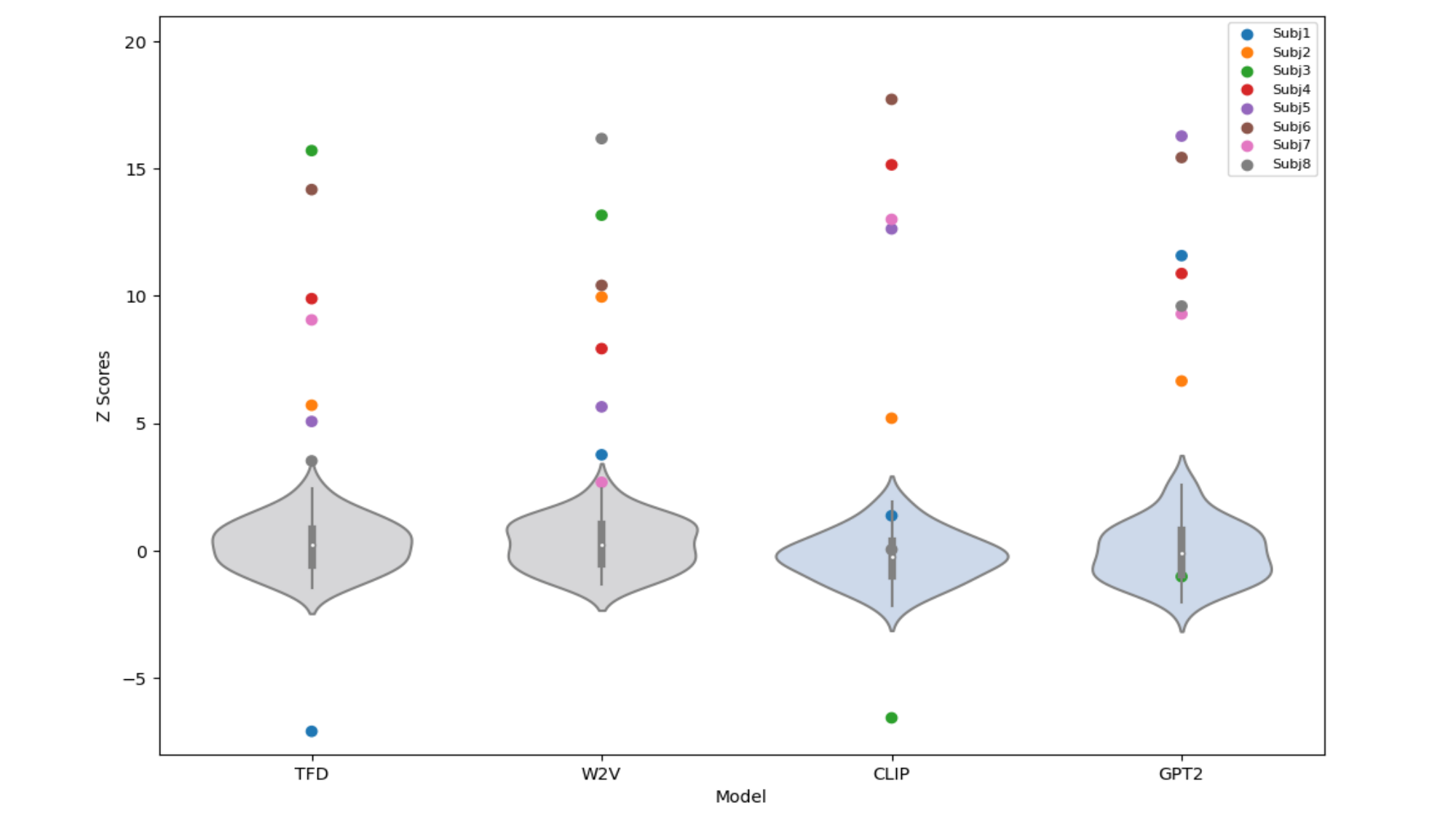}
    \caption{Violin plots depicting the distributions of Z-transformed PC values obtained by regressing real against predicted TFDs for 4 different encoding models (TFD, wav2vec2, CLIP, GPT-2).}
    \label{fig:violinplot}
\end{figure}

Figures~\ref{fig:topomapsz} and~\ref{fig:violinplot} show subject-wise and channel-wise results (respectively) of the empirical test for randomness which our results were subjected to. Figure~\ref{fig:topomapsz} confirms that in all regions where we found the highest reconstruction performances, the latter performances were not random or due to spurious effects.
Z-scores after averaging over channels and subjects are reported in table Table~\ref{tab:combinedzscorepval}. The mean P-values column led to the rejection of the null hypothesis for every encoder, confirming that in all regions where we found the highest reconstruction performances, the latter performances were not random or due to spurious effects. 
Furthermore, we observed Z scores ranged up to 22 and were frequently around the value of 7, pointing towards a high likelihood that the observed effects are significant.

\begin{table}[ht]
\centering
\caption{Summary of statistical validation results for the complete frequency range (0-30 Hz) across different models. The table presents mean (across subjects and channels) P-values and Z-scores for each encoder. The mean P-values represent the probability of observing the data under the null hypothesis of no association between real and predicted TFDs. Significance levels are indicated as percentages, with all P-values being less than the 5\% threshold, denoting statistical significance.}
\label{tab:combinedzscorepval}
\vspace{6pt}
\begin{tabular}{lcccc}
\hline
\multirow{2}{*}{\textbf{Band}} & \multirow{2}{*}{\textbf{Input}} & \multirow{2}{*}{\textbf{Model}} & \multicolumn{1}{c}{\textbf{P-Value}} & \multicolumn{1}{c}{\textbf{Z-Score}} \\ \cline{4-5}
& & & \textbf{mean} & \textbf{mean} \\ \hline
\multirow{4}{*}{Complete} & \multirow{2}{*}{Audio} & TFD & $5.2 \times 10^{-3}$ & 6.1 \\
& & wav2vec2 & $3.2 \times 10^{-3}$ & 6.9 \\
& \multirow{2}{*}{Text} & CLIP & $3.6 \times 10^{-2}$ & 5.8 \\
& & GPT-2 & $2.9 \times 10^{-3}$ & 8.1 \\ \hline
\end{tabular}
\end{table}

\section{Discussion}
The results of this study offer compelling evidence that encoding auditory stimuli is not only achievable but also remarkably precise when using suitable computational methodologies and neural data mapping techniques. This unveils opportunities for comprehending the cognitive processing of auditory information and its wide-ranging applications, spanning from therapeutic interventions to cutting-edge brain-computer interfaces.

\subsection{Brain Representation}
In our results several brain regions exhibited higher correlation across different frequency bands for both models. The Complete frequency spectrum (0-30 Hz) shows extensive regions of higher correlation across the brain for both models, with the most intense positive correlations (red areas) in the frontal cortex, superior temporal gyrus, and parietal lobes. These areas are crucial for processing auditory stimuli, with the frontal cortex involved in attention and cognitive control, the superior temporal gyrus in primary auditory processing, and the parietal lobes in integrating sensory information. Negative correlations (blue areas) are sparsely present in the occipital lobes, suggesting less involvement of visual processing regions during auditory tasks. The Text model also shows extensive regions of higher correlation, with positive correlations dominating the frontal cortex and superior temporal gyrus, although with slightly reduced intensity compared to the Audio model. 
In the Delta band (1-4 Hz), the Audio model shows a mix of positive and negative correlations, with positive correlations scattered in the frontal cortex and medial temporal lobe, regions associated with attention and memory processes crucial during auditory tasks. Negative correlations are observed in the parietal areas, indicating reduced engagement of these regions in low-frequency processing during auditory stimuli. The Text model shows a similar mixed pattern of correlations, with less intensity in positive regions, suggesting that delta band activity is less influenced by text-encoded auditory stimuli.
The Theta band (4-8 Hz) reveals significant negative correlations in the frontal regions for both models, particularly in the prefrontal cortex, involved in working memory and executive functions during auditory processing. Positive correlations are observed in the temporal and parietal regions, indicating theta rhythms’ role in auditory information processing and memory integration. The Text model shows reduced but still present positive correlations in the temporal lobe, highlighting the involvement of theta rhythms in cognitive processing of text-encoded auditory stimuli.
For the Alpha band (8-12 Hz), the Audio model shows positive correlations in the occipital and parietal regions, consistent with alpha rhythms’ association with relaxed states and sensory processing. Negative correlations are prominent in the frontal cortex, suggesting active suppression of irrelevant information during auditory processing. The Text model exhibits a similar pattern with pronounced negative correlations in the frontal regions, indicating alpha rhythms are engaged during both auditory and text-encoded auditory processing, with significant involvement of cognitive control regions.
The Beta band (12-30 Hz) in the Audio model displays mixed areas with positive correlations in the frontal cortex and motor areas, linked to active thinking, focus, and motor planning. Scattered negative correlations are present across the brain, indicating variable engagement of different regions during auditory processing. The Text model shows a similar mixed pattern but with less pronounced correlations compared to the Audio model. Positive correlations in the frontal cortex suggest involvement in cognitive and executive functions during text-encoded auditory processing.
Overall, the topomaps reveal that both types of auditory stimuli engage broad and overlapping brain regions, with distinct patterns of correlation across frequency bands. The frontal cortex, including the prefrontal and motor areas, shows high positive correlations across multiple frequency bands, indicating its critical role in attention, executive functions, and motor planning during auditory processing. The superior temporal gyrus and medial temporal lobe exhibit significant correlations, emphasizing their importance in primary auditory processing and memory integration.
Differences between the Audio and Text models highlight the specialized processing pathways for audio encoded directly versus audio encoded into a latent space using a foundation model. The direct audio model shows more intense correlations in primary auditory regions and frontal areas, reflecting more direct sensory processing. In contrast, the text-encoded model shows broader but less intense correlations, suggesting additional cognitive layers involved in processing encoded auditory information.
The varying intensity and distribution of correlations in frontal, temporal, and parietal regions underscore the differentiated neural engagement required for processing these two types of auditory stimuli. These findings provide insights into the neural mechanisms underlying sensory processing and cognitive functions related to direct and latent space-encoded auditory information, highlighting the complex interplay between sensory input, memory, and executive control in the brain's response to auditory stimuli.

\subsection{Conclusions and Future Directions}
In general, the use of advanced machine learning models, such as transformer-based architectures like GPT or speech-to-text frameworks like wav2vec, while considering their limitations and biases, will remain a key focus in our efforts to unravel the complexities of neural language processing and its applications.
The application of larger, audio or text pre-trained models may influence the outcomes of neural representations, starting from more brain-like features \citep{antonello_scaling_2023}, suggesting a potential area for refinement in future studies. The inclusion of a broader range of subjects and the integration of multimodal data represent exciting avenues for future research. Such expansions would not only enhance the robustness of our findings but also pave the way for a more nuanced understanding of neural processes. 
The potential application of our findings in predicting time series data within neuroscience research introduces new opportunities for advancing the field. Elaborating on the clinical applications of the research findings, such as in diagnosing language disorders or designing neurofeedback interventions, would underscore the translational significance of the study. As bidirectional brain-model mappings grow increasingly powerful, ethical concerns, especially around privacy and misuse, become crucial in neural data studies. It is essential to handle encoding and decoding carefully to prevent biases and protect personal thoughts, underscoring the need for strict ethical guidelines to ensure responsible and privacy-conscious neural research advancements.

\section{Acknowledgements}
This work is supported by NEXTGENERATIONEU (NGEU); the Ministry of University and Research (MUR); the National Recovery and Resilience Plan (NRRP); project MNESYS (PE0000006, to NT) - A Multiscale integrated approach to the study of the nervous system in health and disease (DN. 1553 11.10.2022); the MUR-PNRR M4C2I1.3 PE6 project PE00000019 Heal Italia (to NT); the NATIONAL CENTRE FOR HPC, BIG DATA AND QUANTUM COMPUTING, within the spoke “Multiscale Modeling and Engineering Applications” (to NT); the European Innovation Council (Project CROSSBRAIN - Grant Agreement 101070908, Project BRAINSTORM - Grant Agreement 101099355); the Horizon 2020 research and innovation Programme (Project EXPERIENCE - Grant Agreement 101017727).

\bibliography{neurips_2024}
\bibliographystyle{plainnat}

\end{document}